\documentclass[12pt,citesort]{article}
\usepackage{geometry}
\usepackage{axodraw}
\newcommand{\ice}[1]{\relax}
\usepackage{graphicx}
\usepackage{epsfig}
\usepackage{latexsym}
\usepackage{amsmath}
\usepackage{amsfonts}
\usepackage{amsxtra}
\usepackage{cite} 
\newcommand{\emu}{e_{\mu}}

\newcommand{\Break}{ \nonumber \\ & & }
\newcommand{\smsbar}{\overline{\mbox{{\scriptsize \sc ms}}}} 
\newcommand{\tr}{\text{Tr}}

\def\krto{ {\,\,\lower .8ex\hbox {$\longrightarrow \atop k \rightarrow 0$}\,\,}}
\def\Section#1{\section{#1}\hspace{\parindent}}
\def\subSection#1{\subsection{#1}\hspace{\parindent}}
\def\subsubSection#1{\subsubsection{#1}\hspace{\parindent}}

\def\bea{\begin{eqnarray} }
\def\Ral{{\bf R}$_\alpha\,$}
\def\eea{\end{eqnarray}}

\newcommand{\alvp}{(\frac{\alpha_s}{4\pi})}
\newcommand{\alvps}{\frac{\alpha_s}{4\pi}}
\begin{document}\date{}
\title{The Infrared Behaviour of the Pure Yang-Mills Green Functions}
\author{ Ph.~Boucaud$^a$, J.P.~Leroy$^a$, A.~Le~Yaouanc$^a$, A.Y.~Lokhov$^b$,\\
J. Micheli$^a$, O. P\`ene$^a$, J.~Rodr\'iguez-Quintero$^c$ and 
C.~Roiesnel$^b$ }
\maketitle
\begin{center}
$^a$Laboratoire de Physique Th\'eorique et Hautes
Energies\footnote{Unit\'e Mixte de Recherche 8627 du Centre National de 
la Recherche Scientifique}\\
{Universit\'e de Paris XI, B\^atiment 211, 91405 Orsay Cedex,
France}\\
$^b$ Centre de Physique Th\'eorique\footnote{
Unit\'e Mixte de Recherche 7644 du Centre National de 
la Recherche Scientifique\\ 
}de l'Ecole Polytechnique\\
F91128 Palaiseau cedex, France\\ 
$^c$ Dpto. F\'isica Aplicada, Fac. Ciencias Experimentales,\\
Universidad de Huelva, 21071 Huelva, Spain.
\end{center}
\newcommand{\ghostSD}{\begin{picture}(150,25)(0,0)
\SetWidth{1.2}
\DashArrowLine(12.5,0)(37.5,0){5}
\DashArrowLine(37.5,0)(75,0){5}
\DashLine(75,0)(112.5,0){5}
\DashArrowLine(112.5,0)(137.5,0){5}
\SetWidth{1}
\Vertex(112.5,0){2}
\GlueArc(75,0)(37.5,0,90){-4}{6}
\GlueArc(75,0)(37.5,90,180){-4}{6}
\CCirc(75,0){5}{Black}{Yellow}
\CCirc(75,37.5){5}{Black}{Yellow}
\CCirc(37.5,0){5}{Black}{Yellow}
\Text(20,-10)[l]{a,k}
\Text(50,15)[l]{d,$\nu$}
\Text(100,-10)[l]{e}
\Text(100,15)[r]{f,$\mu$}
\Text(50,-10)[l]{c,q}
\Text(120,-10)[l]{b,k}
\Text(75,48)[c]{q-k}
\end{picture}}
\newcommand{\ghostDr}{\begin{picture}(100,25)(0,0)
\SetWidth{1.2}
\DashArrowLine(12.5,0)(50,0){5}
\DashArrowLine(50,0)(87.5,0){5}
\CCirc(50,0){5}{Black}{Yellow}
\Text(12.5,-10)[l]{a}
\Text(87.5,-10)[r]{b}
\Text(50,-10)[c]{k}
\end{picture}}
\newcommand{\ghostBr}{\begin{picture}(100,25)(0,0)
\SetWidth{1.2}
\DashArrowLine(12.5,0)(87.5,0){5}
\Text(12.5,-10)[l]{a}
\Text(87.5,-10)[r]{b}
\Text(50,-10)[c]{k}
\end{picture}}
\maketitle
\begin{abstract}We study the infrared behaviour of the pure Yang-Mills correlators using
relations that are well defined in the  non-perturbative domain. These are the
Slavnov-Taylor identity for three-gluon vertex and  the Schwinger-Dyson
equation for ghost propagator in the Landau gauge. We also use several inputs
from lattice simulations. We show that lattice data are in serious conflict
with a widely spread analytical relation between the gluon and ghost infrared
critical exponeessencnts. We conjecture that this is explained by a singular
behaviour of the ghost-ghost-gluon vertex function in the infrared. We show
that, anyhow, this discrepancy is not due to some lattice artefact since
lattice Green functions satisfy the ghost propagator Schwinger-Dyson equation.
We also report on a puzzle concerning the infrared  gluon propagator: lattice
data seem to favor a constant non vanishing  zero momentum gluon propagator,
while the Slavnov-Taylor identity (complemented with some regularity hypothesis
of scalar functions) implies that it should diverge.

\end{abstract}
{\begin{flushright}
{\small UHU-FP/05-12}\\
{\small CPHT RR 038.0605}\\
{\small LPT-Orsay/05-38}\\
\end{flushright}
\vfill\newpage

\Section{Introduction}

\subSection{Generalities}

The whole set of correlation functions fully describes a Quantum Field Theory,
as it is related to the S-matrix elements. In  QCD or pure Yang-Mills theories
 Green functions are most often gauge dependent quantities which have 
no direct relationship with physical observables, the latter being necessarily
gauge invariant. However, their indirect physical relevance is well known.
In particular, long distance (or small momentum) Green functions will hopefully
shed some light on the deepest mysteries of QCD such as confinement, spontaneous
chiral symmetry breaking, etc. In this paper we  concentrate our efforts on the study of the  gluon
and ghost Green functions at small momentum in a pure Yang-Mills theory. Our tools
will be Slavnov-Taylor (ST) identities, Schwinger-Dyson
(SD) equations and also several inputs from lattice QCD. 

The infrared behaviour of Green functions has been extensively studied
using different techniques, such as
Schwinger-Dyson equations (see {\it e.g.\/} \cite{Alkofer:2000wg,Fischer:2003rp,Alkofer:2003jj,Bloch:2003yu}
and references therein),
renormalization group methods \cite{Pawlowski:2003hq}, stochastic quantization
\cite{Zwanziger:2003cf,Zwanziger:2001kw}.
 These equations are exact consequences of QCD  and can be easily derived 
 using the path integral formalism. However their practical use reveals in most cases 
 very difficult and one has to resort to a truncation which 
 lessens the rigour of the method. One of the noticeable exceptions is the 
  Schwinger-Dyson equation for the ghost propagator which contains only one 
  integral and thus needs no truncation; this is the only one  which we shall use  in what follows. On the other hand, in order to exploit it  in practice,   one usually has to make appropriate {\sl ans\"atze} for the gluon propagator and the ghost-ghost-gluon vertex. 
 
 We shall also  use the Slavnov-Taylor identity which relates the
three-gluons vertex to the  ghost-ghost-gluon vertex in covariant gauges. Applying
 this relation in the non-perturbative domain will lead, under some 
 assumptions about the infrared regularity of the dressing functions,  to
 non trivial and surprising conclusions. 

 Lattice simulations are, of course, another major tool to study 
 small momentum Green functions. However this paper is not meant to be a standard 
 ``lattice paper''. We aim at using SD and ST to derive properties 
 of the small momentum Green functions and we will use lattice simulations as
 valuable inputs in our theoretical discussion and as a check of some 
hypotheses. As we shall see, the outcome  proves to be quite 
surprising and  undermines some widely spread beliefs.    

In what follows we work in the Landau gauge, but some of the results we present are actually valid in any covariant gauge. Our notations 
are the following :  
\bea
(F^{(2)})^{ab}(k) &=& -\delta^{ab} \frac{F(k^2)}{k^2} \label{dressghost}\\
(G^{(2)}_{\mu\nu})^{ab}(k) &=& \delta^{ab} \frac{G(k^2)}{k^2} \left(\delta_{\mu\nu} -
\frac{k_\mu k_\nu}{k^2} \right)\label{dressglue} \\
\Gamma^{abc}_{\mu\nu\rho}(p,q,r) &=& f^{abc} \Gamma_{\mu\nu\rho}(p,q,r) 
\eea
\bea
\Gamma_{\mu}^{abc}(p,k;q) &\quad =& f^{abc} (-i p_\nu) g_0
\widetilde{\Gamma}_{\nu\mu}(p,k;q) \label{vertghost}  \\
&=& f^{abc} (-i p_\nu) g_0 \nonumber \\ 
& & \cdot  \left[
\delta_{\nu\mu} a(p,k;q) - q_\nu k_\mu b(p,k;q) + p_\nu q_\mu c(p,k;q) 
 + q_\nu p_\mu d(p,k;q)  \right. \nonumber\\ & &\quad + \left.  p_\nu p_\mu e(p,k;q)\right] \nonumber 
\label{vertgluon}
\eea
respectively for the ghost propagator, the gluon propagator, the three-gluons vertex and the
ghost-ghost-gluon vertex\footnote{We stick to the decomposition given in ref.  \cite{Chetyrkin:2000dq} {\sl except for  the arguments of the scalar functions, for which we keep the same order as in $\Gamma$ itself }}. All momenta 
are taken as entering. In eq.~(\ref{vertghost}) 
 $-p$ is the momentum of the outgoing ghost, $k$ 
 the momentum of the incoming one and $q=-p-k$ the momentum of the gluon.
  $F(p^2)$ and $G(p^2)$ are the dressing   functions
of the ghost and  gluon propagators respectively.
We parameterise the propagators in the infrared by setting at leading order
\begin{equation}
\label{param}
\begin{split}
& G(p^2)=\left(\frac{p^2}{\lambda^2}\right)^{\alpha_G}  
\\ & F(p^2)=\left(\frac{p^2}{\eta^2}\right)^{\alpha_F}, \quad\text{when }p^2\text{ is small},
\end{split}
\end{equation}
where $\lambda,\eta$ are some dimensional parameters. 

Let us make a brief and partial summary of the present predictions for $\alpha_{F,G}$ (\cite{Alkofer:2000wg,Fischer:2003rp,Alkofer:2003jj,Bloch:2003yu,vonSmekal:1997is,Lerche:2002ep,Pawlowski:2003hq,Zwanziger:2001kw}). We refer, for a very complete list of references, to the review by Alkofer and von Smekal \cite{Alkofer:2000wg}.  All those references assume  the relation $2 \alpha_{F} + \alpha_{G} = 0$ and parameterise $\alpha_{F}$ and $\alpha_{G}$  as

 \begin{equation}\begin{array}{l}
\alpha_F=-\kappa_{\text{SD}}\\
\alpha_G=2 \kappa_{\text{SD}}.\end{array}
\end{equation}


Different truncation schemes for the 
Schwinger-Dyson equations give (\cite{Alkofer:2000wg,Fischer:2003rp,Alkofer:2003jj,Bloch:2003yu,vonSmekal:1997is,Lerche:2002ep,Pawlowski:2003hq})

\begin{equation}
\begin{array}{ll}\kappa_{\text{SD}} =0 .92 & \mbox{\cite{vonSmekal:1997is}}\nonumber\\
\kappa_{\text{SD}} \in [0.17,0.53]&\mbox{ \cite{Bloch:2003yu} , using a 2-loop perturbative input}\nonumber
\end{array}\end{equation} 
while another approach, (\cite{Zwanziger:2001kw, Pawlowski:2003hq}),  predicts two possible solutions
\begin{equation}
\begin{array}{l}
\kappa_{\text{SD}}=1,\quad\text{or} \\
\kappa_{\text{SD}} =0.59\nonumber
\end{array}
\end{equation}
Lattice simulations give  $\alpha_G \simeq 1$. Note   that $\alpha_G= 1$ corresponds  to 
the gluon propagator being finite and non-zero at vanishing momentum; in other words, among the numbers we have just quoted, only the ones given in ref.~\cite{Bloch:2003yu} lead to a divergent infrared gluon propagator while the other values correspond to a vanishing one\footnote{After the completion of this paper our attention was drawn on ref.~\cite{Kato:2004ry} which also predicts an IR-divergent gluon propagator and on refs.~\cite{Kondo:2003sw, Kondo:2003rj} which lead to a finite non-zero one.}.

\subSection{Numerical setup of the lattice simulations}
\label{intro2}
In this part we briefly describe the technical details of  our numerical lattice simulations.

We  use the standard Wilson action. For the 
$SU(2)$ gauge group we have  used  lattices  of size  $32^4$   and  $48^4$ with
$\beta=2.3$. This value of $\beta$ corresponds to 
$\beta_{\text{SU(3)}}\approx 5.75$. In the case of the $SU(3)$ gauge group the
simulation has been done on  $32^4$,  $24^4$ and smaller  lattices with $\beta = 5.75$ and $\beta = 6.0$. Those rather low values of $\beta$ have been chosen because they allow measurements at small momenta. We have used
periodical boundary conditions for the gauge field.  The gluon propagator is defined as
a mean value over gauge field configurations  :

$$
< A^a_\mu(x) A^b_\nu(y)>.
$$
The ghost propagator is calculated by the inversion of the discretised 
Faddeev-Popov operator  (cf eq.~\ref{FP_lattice}).
For this purpose  we have used the conjugate gradient algorithm with the source 
$$
 \left(1-\frac{1}{V}, \frac{1}{V} , \ldots, \frac{1}{V}\right)
$$
where $V$ is the number of lattice points.
The r\^ole of the $1/V$ terms is  to eliminate the zero modes
of the Faddeev-Popov operator (corresponding to global 
gauge transformations) in order to allow its inversion in the orthogonal subspace (cf ref.~\cite{Claude:FP}). All  lattice data  have been extrapolated to the continuum as described  in ref.~(\cite{Claude:FP}).
A detailed report on all  
numerical results is presented in the same reference.


\Section{Constraints on $\alpha_F$ and $\alpha_G$ from the ghost \\Schwinger - Dyson equation}

There is a widely used relation between $\alpha_F$ and $\alpha_G$, referred to  in the following as \Ral, which comes  from the scaling 
analysis  of the Schwinger-Dyson equation for the ghost propagator (\cite{Zwanziger:2001kw}%
) and states that in
four-dimensional space one has
\label{alphas_relation}
$2\alpha_F+\alpha_G=0.$

We have  attempted to test this relation on lattices  with the characteristics indicated above. We plot in fig.~\ref{Both_F2G}
the quantity $F^2(p) G(p)$ as a function of $p$.  If the relation \Ral was true,  this quantity should be constant in the infrared
domain. One can see   that {\it it is not the case: $F^2G$
goes to zero at small momenta}. On the other hand  the ultraviolet (UV) behaviour is exactly the expected one.
The  same trend is already visible at  $\beta$'s larger than ours : in
refs.~(\cite{Nakajima:2004vc, Furui:2004bq}) it is mentioned that $F^2G$ might decrease
as the momentum approaches zero\footnote {In ref. \cite{Nakajima:2004vc}
it is assumed that the vertex function stays constant in the zero momentum
 limit, in which case $F^2G$ is proportional to the strong coupling constant 
 $\tilde \alpha_s$ in the MOM scheme based on ghost-ghost-gluon vertex.} at
 $\beta=6.0$ and $\beta=6.4$.
 The same authors  have even reported on  the same effect in the unquenched case 
 \cite{Furui:2004uy}. Very recently  Ilgenfritz et al. have published in ref.~\cite{ilgenfritz} results which go in the same direction (although the conclusions they draw 
  thereof differ  from ours).

It was suggested in~\cite{Furui:2004bq} that Gribov copies might induce  significant changes
in the  infrared behaviour of the ghost propagator. Could this explain our 
findings for $F^2 G$~? In  ref.~ \cite{Bakeev:2003rr}  the accurate  lattice
gauge fixing (choosing the "best copy", corresponding to the lowest value 
of $\| A ^{(g)}\|$)  seems to
lessen the infrared divergence of the ghost propagator \cite{Bakeev:2003rr, ilgenfritz2, IlgenGrib}, implying a further increase of the drop of $F^2G$ in the infrared, 
while the gluon propagator
is known to be only slightly affected by the presence of lattice Gribov copies. 
Furthermore, we shall see that the SD equation is closely related to a 
lattice-SD equation which is a mathematical identity, valid independently 
of the choice of the Gribov copy. Thus the
possible influence of Gribov copies on propagators cannot explain the behaviour
of $F^2 G$ at small momenta.
%

\begin{figure}[h]
\hspace{\baselineskip}
\begin{center}
\hspace*{-.5cm}\begin{tabular}{lcr}
\includegraphics[width=7.5cm]{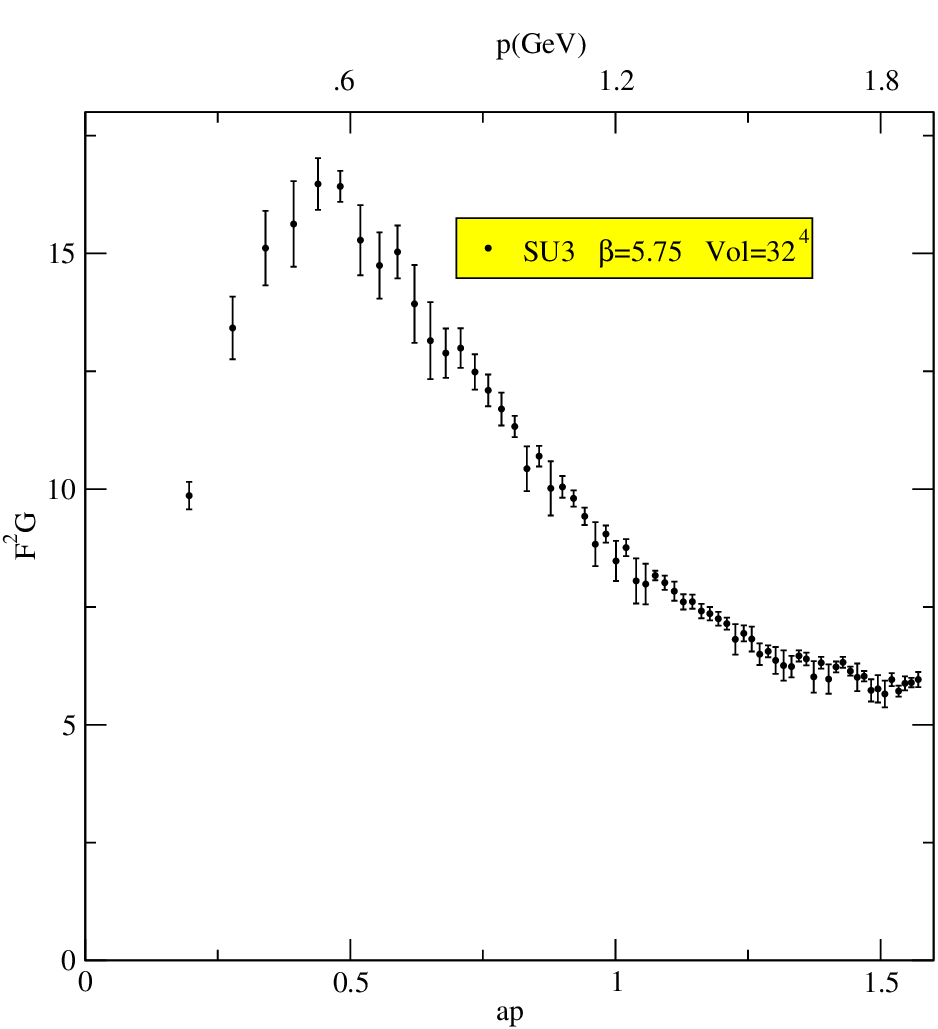}&&\includegraphics[width=7.5cm]{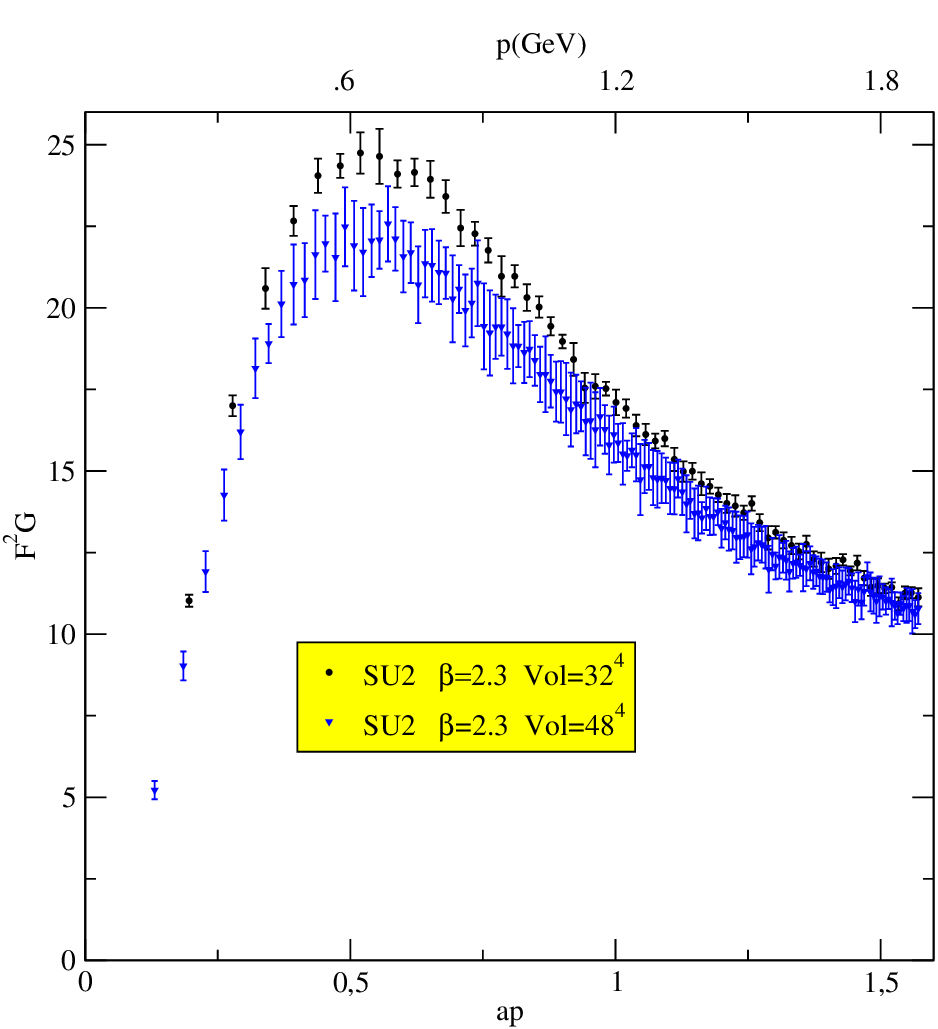}
\end{tabular}
\end{center}
\caption{$F^2 G$ from lattice simulation for the $SU(3)$ (left, $32^4, \beta=5.75$)  and $SU(2)$
(right, $32^4 $ and $48^4,\beta_{SU(2)}=2.3$) gauge groups. The $\beta$'s are chosen so as to give the same lattice spacings : 1.2 $GeV^{-1}$.  If the relation $2\alpha_F + \alpha_G = 0$ was true this quantity should be constant in the infrared domain. One clearly does not see this behaviour.}
\label{Both_F2G}
\end{figure}
The rest of this section is aimed at understanding this disagreement 
between lattice simulations and the theoretical claim (\Ral). 
We will first revisit the proof of the latter in order
to identify all the hypotheses needed and submit each of them to a
critical analysis. 
 In the next-to-following section we will discuss a
special writing of the ghost Schwinger-Dyson equation, in a form which involves only Green
functions instead of vertices and can thus be directly tested on the
lattice.

\subSection{Revisiting the relation between $\alpha_F$ and $\alpha_G$ 
}
\label{revisiting}
We will examine to what extent the proof of \Ral  is compelling, using the Schwinger-Dyson equation for the bare ghost
propagator  which can be written diagrammatically as

\vspace{\baselineskip}
\begin{small}
\bea
\left(\ghostDr\right)^{-1}%
=
\left(\ghostBr\right)^{-1}%
- 
\ghostSD %
\nonumber
\eea\end{small}%
\noindent i.e.

\bea
\label{SD}
(F^{(2)})^{-1}_{ab}(k) &=&-\delta_{ab} k^2  \\ 
& &- g_0^2 f_{acd} f_{ebf} 
\int \frac{d^4q}{(2\pi^4) }  F^{(2)}_{ce}(q)
(i q_{\nu'}) \Gamma_{\nu'\nu}(-q,k;q-k) (i k_\mu) (G^{(2)})_{\mu\nu}^{fd}(q-k), \nonumber
\eea

where we use the notations of eqs. (\ref{dressghost}-\ref{vertgluon})
 and $g_0$ is the bare coupling constant. This integral equation is written in terms of  
 bare Green functions. It can be cast into a renormalized form
 by  multiplying $G^{(2)}$ (resp. $F^{(2)}, \Gamma$) by $Z_3^{-1}$ (resp. $\widetilde Z_3^{-1}, \widetilde Z_1$) and
 $g_0^2$ by $Z_g^{-2} = Z_3 \frac{ \widetilde Z_3^2}{ \widetilde Z_1^2}$ and multiplying  the $k^2$ term by
 $\widetilde Z_3$. The integral therein is ultraviolet divergent 
 but one can check that the cut-off dependence is matched by the cut-off 
 dependence of $Z_g$ and of the $\widetilde Z_3$ factor multiplying $k^2$. Later on we will only use
\emph{subtracted} SD equations such that the UV divergence is cancelled as well as the
 $\widetilde Z_3 k^2$ term. These subtracted SD equations hold both in terms 
 of bare and renormalized Green functions without any explicit renormalization
 factor.  
Let us now consider (\ref{SD}) at small momenta $k$.
 The ghost-gluon vertex may be expressed 
as
\bea
q_{\nu'} \widetilde{\Gamma}_{\nu'\nu}(-q,k;q-k) =  q_\nu H_1(q,k) + (q-k)_\nu H_2(q,k)
\eea
where,  using the decomposition (\ref{vertghost}), 
 one gets: 

\bea\label{h1h2}
H_1(q,k) &=& a(-q,k;q-k) - (q^2 -q{\cdot}k) \left( b(-q,k;q-k)  \right. \nonumber\\  & &+ \left.  d(-q,k;q-k)\right) + q^2  e(-q,k;q-k) )\simeq \nonumber \\
 &\stackrel{{k\to 0}}{\simeq}& a(-q,k;q-k) - q^2 \left( b(-q,k;q-k) + d(-q,k;q-k) - e(-q,k;q-k) \right) \nonumber \\
 H_2(q,k) &=& (q^2 -q{\cdot}k)\,  b(-q,k;q-k) - q^2\, c(-q,k;q-k) %
\eea

In the Landau gauge, because of the transversality condition, $H_2$ does not contribute. Thus, dividing both sides of eq.~(\ref{SD})  by $k^2$ and omitting colour indices, one obtains
\begin{equation}
\label{SD1}
\begin{split}
\frac{1}{F(k)} & = 1 + g_0^2 N_c \int \frac{d^4 q}{(2\pi)^4} 
\left( \rule[0cm]{0cm}{0.8cm}
\frac{F(q^2)G((q-k)^2)}{q^2 (q-k)^2} 
\left[ \rule[0cm]{0cm}{0.6cm}
\frac{\frac{(k\cdot q)^2}{k^2} - q^2}{(q-k)^2}  
        \right]
\ H_1(q,k)
           \right).
\end{split}
\end{equation} 
\subsection{What does the  ``non-renormalization theorem" exactly say?}

A widely used statement, known as the ``non-renormalization theorem", claims that, in the Landau gauge, the renormalization constant $\tilde Z_1$ of the ghost gluon vertex is exactly one. Note that there is no reference to a particular 
renormalisation scheme. Formulated in this way, this claim is wrong. Let us first state
and then explain below what is true in our opinion :

1) There is a true and very clear statement which can be extracted from Taylor's paper (the argument
is given below), ref.~\cite{Taylor}.
\bea
{\widetilde{\Gamma_\mu^{abc,Bare}}(-p,0;p)}= -i f^{abc}p_\mu \label{taylor}
\eea
i.e. there is no radiative correction in this particular momentum configuration
(with zero momentum of the ingoing ghost)

2) This entails that ${\widetilde{\Gamma}_\mu^{abc,Bare}(p,k;q)}$ is {\it finite} whatever the external momenta, and that therefore $\widetilde{Z_1}^{\overline{\mbox{\sc ms}}}=1$. In addition, we get also trivially $\tilde{Z_1}^{{\mbox{\scriptsize \sc MOM}_h}}=1$, where {\sc MOM}$_h$ refers to the configuration of momenta in equation~(\ref{taylor}).
In general, in other schemes, {\it there is} a finite renormalisation, and this is why
we do not adopt the misleading expression "non-renormalization theorem".

3) In particular, one finds in the very extensive calculations of radiative corrections at least two cases of {\sc MOM} schemes where there is a finite renormalisation (and certainly many more) : {\sc MOM}$_g$ in the notations of  ref.\cite{Chetyrkin:2000dq}, and the {\it symmetric } {\sc MOM} scheme. For the latter, we give the proof below.

The essence of Taylor's argument  is actually very simple. In a kinematical situation where the incoming ghost momentum is zero, consider any perturbative contribution to the ghost-gluon vertex. Following the ghost line in the direction of the flow,  the first vertex will be proportional to the outgoing ghost momentum $p_\mu$, i.e. to  the gluon momentum  $-p_\mu$.  In the Landau gauge this contribution will thus give 0 upon contraction with the gluon propagator $D_{\mu\nu}(p)$. Therefore the only contribution to  remain is  the tree-level one.  In other words the bare ghost-gluon vertex  is shown to be  equal to its tree-level value in these kinematics : ${\widetilde{\Gamma_\mu^{abc,Bare}}(-p,0;p)}= -i f^{abc}p_\mu$. 
This  result has been checked by means of  a direct evaluation  to  three loops in perturbation theory by Chetyrkin. 
In our notations  :
$$H_1(p,0) + H_2(p,0)=1.$$  
Note that in the Schwinger-Dyson equation \ref{SD1}, only
$H_1$ is present, and the theorem of Taylor does not tell that $H_1(p,0)=1$,
as seems assumed in many Schwinger-Dyson calculations, where it plays a crucial r\^ole in the proof  of \Ral.

\begin{figure}[h]
\begin{center}
\psfig{figure=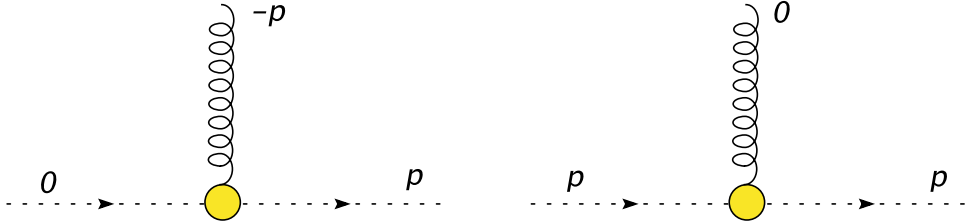,height=3cm}
\end{center}
\caption{The kinematical situations considered below. The left diagram (0-momentum incoming ghost) corresponds to $\Gamma_h$ below which is known to be equal to one. The right one (0-momentum gluon) corresponds to $\Gamma_g $ and  leads to a non-trivial $p^2$-dependence}
\label{cinem}
\end{figure}

As an illustration of our point 3), let us quote the formulas from the appendix of ref.~\cite{Chetyrkin:2000dq}, reduced to the situation we are interested in ($\xi_L = 0, n_f = 0$). The two dressing functions $\tilde\Gamma_h$ (resp. $\tilde\Gamma_g$) are defined by $\widetilde \Gamma_\mu^{abc}(-p,0,p)= -i  f^{abc}\tilde\Gamma_h(p)$  (resp. $\widetilde \Gamma_\mu^{abc}(-p,p,0)= -i  f^{abc}\tilde\Gamma_g(p)$ ) and correspond to the kinematical situations depicted in the left (resp. right) part of fig.~(\ref{cinem}). We have already mentionned that $\Gamma_h$ is exactly one, but this does not hold for $\Gamma_g$ and, indeed, one has at three loops :
 \begin{eqnarray}
\label{GammagrenMSb}
\tilde \Gamma_{\mathrm{g}}^{\smsbar}\vert_{p^2=\mu^2} & = & 
1\;
{}
+ \,\frac{3}{4}\,  \alvps\, C_A\;
+ \, \frac{599}{96}    \alvp^{2}     C_A^{2}\;
+ \,   \left[
     \frac{43273}{432} 
    + \frac{783}{64}      \zeta_3
    - \frac{875}{64}      \zeta_5
\right]  \alvp^{3}     C_A^{3}\;\Break
+ \, \left[
     \frac{27}{4} 
    - \frac{639}{16}      \zeta_3
    + \frac{225}{8}      \zeta_5
\right]    \alvp^{3}     C_A^{2}     C_F\;
{}.\end{eqnarray}

It is then easy to find the  $p^2$-dependence  :

\begin{eqnarray}
\tilde\Gamma_{\mathrm{g}}=\widetilde \Gamma_{\mathrm{g}}^{\smsbar}\vert_{p^2=\mu^2} + \left[ \frac{11}{4} C_A^{2} \alvp^{2} +\frac{7813}{144} C_A^{3} \alvp^{3} + \cdots\right] \log(\frac{\mu^2}{-p^2}) +\cdots\nonumber
\end{eqnarray}

In ref.~\cite{Lerche:2002ep} the non -renormalization theorem is understood as the statement that the vertex reduces to its tree-level form at all symmetric-momenta points in a symmetric subtraction scheme. However this statement is not supported by a direct  evaluation. Using the one-loop results of Davydychev (ref.~\cite{Davydychev:1996pb}) one gets in  a symmetric configuration the value 

\begin{eqnarray}
\widetilde \Gamma_\mu^{abc}(p,k;q)\vert_{p^2=k^2=q^2=\mu^2}= -i  f^{abc}\left\{p_\mu \,\left(1 +\,\alvps\, \frac{C_A}{12}(9+ \frac{5}{2}\phi) \right)+q_\mu\, \alvps\, \frac{C_A}{12}(3+ \frac{5}{4}\phi)\right\}
\end{eqnarray} 
with $\phi=\frac{4}{\sqrt{3}}Cl_2(\frac{\pi}{3}), Cl_2(\frac{\pi}{3})=1.049\cdots$. 

According to   ref.~\cite{Lerche:2002ep} the coefficient  of $p_\mu$ should be one. The presence of $\alpha_s$  in the above formulas implies on the contrary that the vertex will in general depend on the momenta  : using the results given in the appendices of ref.\cite{Chetyrkin:2000dq} one finds for the  leading $p^2$-dependence 
$ -i  f^{abc}\left\{\frac{11}{3}\frac{C_A^2}{12}\alvp^2\log(\frac{p^2}{\mu2})\left( (9+\frac{5}{2}\phi) p_\mu + (3+ \frac{5}{4}\phi) q_\mu\right)\right\}$. This dependence is   logarithmic, as is expected   in a perturbative approach.   Furthermore, in ref.~(\cite{Lerche:2002ep}) it is supposed that the vertex function takes the form $(q^2)^\ell (k^2)^m ((q-k)^2)^n$ with the restriction $\ell+m+n=0$. One should note that this last condition corresponds in our notations to $\alpha_\Gamma = 0$ (cf. section ~(\ref{subtracted}) below). This restriction comes from the assumption that the {\emph symmetric} vertex is equal to 1 for any $p^2$, which, as we have just seen, is actually not the case. 
Therefore  we  shall adopt a more general point of view and keep open the possibility of a 
non perturbative effect leading to a singular or  vanishing  limit of $H_1$ when
$q \to 0$. We should mention that the problem of the $p^2$-dependence of the ghost-gluon  vertex has already been addressed in refs.~\cite{Alkofer:2004it, Schleifenbaum:2004id}. However these authors work under the condition  \Ral which appears not to be 
satisfied  by our lattice data.

\subsection{A subtracted Schwinger-Dyson equation} \label{subtracted}
Let us now consider two infrared scales $k_1\equiv k$ and $k_2\equiv \kappa
k$.  Calculating the difference of eq.~(\ref{SD1}) taken at scales $k_1$
and $k_2$ and supposing for the moment that $\alpha_F\neq 0$ one obtains
\begin{equation}
\label{k1k2}
\begin{split}
\frac{1}{F(k)} - \frac{1}{F(\kappa k)} & \propto
(1-\kappa^{-2\alpha_F} )  (k^{2})^{-\alpha_F}
=  g_0^2 N_c \int \frac{d^4 q}{(2\pi)^4} 
\left( \rule[0cm]{0cm}{0.8cm}
\frac{F(q^2)}{q^2} \left(\frac{(k\cdot q)^2}{k^2}-q^2\right) \right.
\\ & \left. \times
\left[ \rule[0cm]{0cm}{0.6cm}
\frac{G((q-k)^2)H_1(q,k)}{\left((q-k)^2\right)^2} -  
\frac{G((q-\kappa k)^2)H_1(q,\kappa k)}{\left((q-\kappa k)^2\right)^2}
\rule[0cm]{0cm}{0.6cm} \right]
\rule[0cm]{0cm}{0.8cm} \right).
\end{split}
\end{equation}
We now  make the hypothesis that there exists a scale $q_0$ such that
$$G(q^2) \ \sim \ (q^2)^{\alpha_G}, \quad
F(q^2) \ \sim \ (q^2)^{\alpha_F}  , \quad \mbox{\rm for}  \quad q^2 \le q_{0}^2. $$\

\noindent Similarly, we suppose that $H_1$ can be written for  $ q^2, k^2 \le q_{0}^2$ as

$$H_1(q,k) \ \sim \ (q^2)^{\alpha_\Gamma} \
 h_1\left(\frac{q\cdot k}{q^2},\frac{k^2}{q^2} \right) $$

\noindent or as

$$H_1(q,k) \ \sim \ ((q-k)^2)^{\alpha_\Gamma} \
 h_2\left(\frac{q\cdot k}{q^2},\frac{k^2}{q^2} \right)$$

\noindent where the scalar functions $h_{1,2}$ are supposed to be regular enough (i.e. free of singularities  worse 
than logarithmic) and expandable in Taylor series for $k \to 0$
They are obviously invariant under any simultaneous rescaling of both 
$q$ and $k$. The exponent $\alpha_\Gamma$ gives the leading critical behaviour of $H_1$ on $q$.



 Thus we  rewrite (\ref{k1k2}) by rescaling $k \to \lambda k$ with $\lambda$ chosen
so that $(\lambda k)^2 \ll q_{0}^2$ and  splitting the integral in the r.h.s.
into two parts 
\begin{equation}
I_1(\lambda)=\int_{q^2 < q_{0}^2} \frac{d^4 q}{(2\pi)^4} \Big[ \ldots \Big], \qquad
I_2(\lambda)=\int_{q^2 > q_{0}^2} \frac{d^4 q}{(2\pi)^4} \Big[ \ldots \Big].
\end{equation}
In $I_1$, since $(\lambda k)^2 \ll q_{0}^2$, we can substitute  the infrared approximations
(\ref{param}) for $G$ and $F$   . $I_1$ is infrared convergent if :
\begin{eqnarray}
\label{cond_ward}
\alpha_F + \alpha_\Gamma &>& -2 \qquad{\rm IR\; convergence \; at}\; q^2 = 0 \nonumber \\
\alpha_G + \alpha_\Gamma&>& -1 \qquad{\rm IR\; convergence\; at}\; (q-k)^2 = 0 \; 
{\rm and} \;(q-\kappa k)^2 = 0
\end{eqnarray}
We shall suppose in the following that these conditions are verified.
We then obtain, performing the change of variable $q\to \lambda q$  and writing generically $h$ for $h_{1,2}$:
\bea
 &I_1(\lambda)\approx \lambda^{2\left(\alpha_F + \alpha_G + \alpha_\Gamma
\right)} \displaystyle \int_{q^2 < 
\frac{q_{0}^2}{\lambda^2}} \frac{d^4 q}{(2\pi)^4} 
\left( \rule[0cm]{0cm}{0.8cm} %
(q^2)^{\alpha_F +\alpha_\Gamma -1}
        \left(
                \frac{(k\cdot q)^2}{k^2}-q^2
        \right) \nonumber \right.
\\
& \left. \times 
        \left[ 
                \left(  (q-k)^2 \right)^{\alpha_G  -2}
		h\left(\frac{q\cdot k}{q^2},\frac{k^2}{q^2} \right) -
                \left((q-\kappa k)^2\right)^{\alpha_G
		 -2} h\left(\kappa\frac{q\cdot k}{q^2},
		 \kappa^2\frac{k^2}{q^2} \right)
        \right]
\rule[0cm]{0cm}{0.8cm} \right) \ . %
\label{I1}
\eea
The point we have to keep in  mind is the fact that
the upper bound of the integral goes to infinity when $\lambda \to 0$. This potentially induces a dependence on $\lambda$ whose interplay with the behaviour explicitly shown in (\ref{I1}) we must check.
In this limit, the convergence of the integral depends on the asymptotic 
behaviour of the whole integrand for large $q$. In particular, the 
leading contribution of the square bracket in  eq.~(\ref{I1}) behaves 
as 

$$
(q^2)^{\alpha_G-2} \left(\frac {k^2}{q^2}\right)  \ \sim \ q^{2\alpha_G-6} \ ,
$$

\noindent because the  terms in $q \cdot k$,  being odd under
 $q_\mu \to -q_\mu$, give a null contribution under
the angular integration in eq.~(\ref{I1}). Thus, assuming  the conditions~(\ref{cond_ward}) are satisfied  the integral $I_1(\lambda)$
is   guaranteed to be convergent when  $q \to 0$ (or $k \to q$) and its asymptotics for small $\lambda$ 
 is given by

\bea
I_1(\lambda) \sim \left\{ 
\begin{array}{ll} 
\displaystyle \lambda^{2(\alpha_G+\alpha_F+\alpha_\Gamma)} \int_0^{q_0/\lambda} dq \ q^{2(\alpha_F+\alpha_G+\alpha_\Gamma)-3} 
\ \sim \ &\lambda^{2(\alpha_G+\alpha_F+\alpha_\Gamma)} \\
 & \mbox{\rm if} \ \ \alpha_G+\alpha_F+\alpha_\Gamma  < 1\\
\displaystyle \lambda^2 \int_0^{q_0} dq \ q^{2(\alpha_F+\alpha_G+\alpha_\Gamma)-3} \ \sim \ \lambda^2 & \mbox{\rm if}   \ \ \alpha_G+\alpha_F+\alpha_\Gamma  \ge 1
\end{array} 
\right.
\label{cond_alpha_F_G}
\eea
because in both cases the integral on the momentum $q$ is finite and does not depend on $\lambda$ in the limit
 $\lambda \to 0$.

Let us now consider  $I_2$. Its  dependence on $\lambda$ is explicit in the factor   
$$ \frac{G\left((q-\lambda k)^2\right) H(q,\lambda k)}{\left((q-\lambda k)^2\right)^2}
-\frac{G\left((q-\lambda \kappa k)^2\right)
H(q,\lambda\kappa k)}{\left((q-\lambda \kappa k)^2\right)^2}$$ which stems from the substitution $k \to \lambda k$ in~(\ref{k1k2}). Clearly, this quantity can only be even in $\lambda$ : any odd power of  $\lambda$  would imply an odd power of $q \cdot k$ whose angular integral is zero.  Since the integrand is identically zero at $\lambda = 0$ and the integral is ultraviolet convergent, it is  proportional to $\lambda^2$ (unless some accidental cancellation forces it to behave as a higher even power of  $\lambda$).

So, if the first of the conditions~(\ref{cond_alpha_F_G}) is verified, it follows  $I_1 +I_2 \approx
\lambda^{2(\alpha_F+\alpha_G+\alpha_\Gamma)}$, else 
$I_1 +I_2 \approx \lambda^2$. Comparing
this to the left hand side of eq.~(\ref{SD1}) 
\begin{equation}
\label{powerlambda}
\frac{1}{F(\lambda k)} - \frac{1}{F(\lambda \kappa k)} \sim \lambda^{-2(\alpha_F )}
\end{equation}
we then conclude :
\begin{eqnarray}\label{choix}
\alpha_F +\alpha_G +\alpha_\Gamma &<1\,&\implies 2\alpha_F + \alpha_G +\alpha_\Gamma 
= 0 \nonumber \\
\alpha_F +\alpha_G +\alpha_\Gamma &\ge1\,&\implies \alpha_F =-1
\end{eqnarray}

In the particular case $\alpha_F = 0$  the leading term in the l.h.s. of eq.~(\ref{k1k2}) is identically zero.  We are left with the subleading one which, pursuing  the same argumentation,  we suppose to be proportional to $k^2$. Then the argument is the same as in the previous case except for the power of $\lambda$ in the r.h.s. of  eq.~(\ref{powerlambda}) which becomes equal to 2.  It results that the case 
$\alpha_G +\alpha_\Gamma < 1$ is now excluded while the case $\alpha_G +\alpha_\Gamma \ge 1$ provides no extra constraint.

\begin{table}\begin{small}
\begin{tabular}{|l|c|c|c|c|}
\hline
Case&$\alpha_F \ne 0$&$\alpha_F \ne 0$&$\alpha_F = 0$&$\alpha_F = 0$\\
&$\alpha_F +\alpha_G +\alpha_\Gamma < 1$&$\alpha_F +\alpha_G +\alpha_\Gamma \ge1$&$\alpha_G +\alpha_\Gamma < 1$&$\alpha_G +\alpha_\Gamma \ge1$\\
\hline
IR &&&&\\
Convergence&$\alpha_F  +\alpha_\Gamma > -2$&$\alpha_F  +\alpha_\Gamma> -2$ &$\alpha_\Gamma> -2$&$  \alpha_\Gamma> -2$\\
Conditions&$\alpha_G +\alpha_\Gamma> -1$&$\alpha_G +\alpha_\Gamma> -1$&$ \alpha_G +\alpha_\Gamma> -1$&$\alpha_G +\alpha_\Gamma> -1$\\
\hline
SD&&&&\\
 constraints& $2\alpha_F + \alpha_G +\alpha_\Gamma 
= 0$& $\alpha_F =-1$&excluded&none\\
\hline
\end{tabular}\caption{Summary of the various cases regarding the $\alpha$ coefficients}\label{tabledescas}
\end{small}\end{table}

The various possibilities which have appeared in this discussion are summarised in table~\ref{tabledescas}. From this table it appears that \emph{only the triple
condition that}\\
 $$\alpha_F \ne 0, \quad  \alpha_\Gamma=0, \quad  \alpha_F +\alpha_G < 1$$ \\ does actually imply the \emph{standard statement that } $2\alpha_F + \alpha_G =0$ .
However, the plot in Fig.\ref{Both_F2G} indicates a behaviour 
$2\alpha_F +\alpha_G > 0$, indicating that at least one of these  conditions 
is not fulfilled. 

Let us  assume for the moment that $\alpha_F +\alpha_G\ge 1 $ and $\alpha_\Gamma=0$.
Then $2\alpha_F +\alpha_G\ge 0$, in agreement with
fig.\ref{Both_F2G}, and  $\alpha_G \ge 2$.  However the possibility that $\alpha_G $ be greater than 2  is unambiguously excluded by  the lattice simulations so that  the hypothesis has to be rejected. Furthermore, 
as we shall see, one can derive from the Slavnov-Taylor identity relating
the  three-gluons and ghost-ghost-gluon vertices the inequality $\alpha_G < 1$
if one assumes that some  of the scalar form factors of these vertices are regular
 when one momentum goes to zero. 

We now consider the hypothesis $2\alpha_F + \alpha_G +\alpha_\Gamma 
= 0 $ with $\alpha_\Gamma< 0 $ to comply with the lattice indications  of 
fig.\ref{Both_F2G}. This  implies 
that some of the scalar factors of the ghost-ghost-gluon vertex are singular in the infrared.
We shall turn back to this possibility in the concluding remarks of this section~(\ref{concluSD}). The question is whether a non-perturbative effect could generate 
a non vanishing $\alpha_\Gamma$. 

A direct lattice estimate of the ghost-ghost-gluon vertex would be welcome.
However this is a difficult task.  A direct measurement implying a zero momentum ghost is 
impossible since the corresponding Green functions are singular because of the zero modes of the Faddeev-Popov operator.  A careful limiting procedure implying very small external  momenta
has to be performed.  This study is under way.
In between we propose a simpler check based on another writing of the SD equation 
(\ref{SD})  in terms of a pinched Green function which can be directly checked on the lattice. 
This will also allow to control the lattice artefacts and to rule out the hypothesis  that the problem
 encountered in fig.~\ref{Both_F2G} is simply due to a lattice artefact.

%
\subSection{Green function formulation of the ghost SD equation }

We will now rewrite the ghost SD equation using only propagators and the
ghost-ghost-gluon Green function. In this form, its validity can be  tested on
the lattice, because what one directly calculates  in lattice
simulations are {\it Green functions},  not  {\it vertex functions}.  
If we consider the loop integral in Eq.~(\ref{SD}) we can see that it is nothing 
else but a ghost-ghost-gluon Green function in which the left ghost leg 
has been cut and  where the gluon and right-hand ghost have been pinched 
onto the same point in configuration space. We shall see that this quantity
is  directly accessible from lattice data \emph{without making any
specific assumption about the behaviour of the vertex function}.

The interest of this approach is that it will help us to throw a closer 
look at the compatibility of the lattice simulations and the SD equations. 
Indeed, as we have just  seen, we are facing a contradiction between, on the one hand the
lattice estimate of $F^2\, G$ (fig. \ref{Both_F2G}) and, on the other hand, the
relation  \Ral which is derived from the SD equation 
(\ref{SD}) complemented by a regularity assumption ($\alpha_\Gamma=0$) 
 suggested by perturbation theory. Therefore we feel the need to directly
confront  lattice calculations with SD. The form of SD which is presented in this subsection 
allows such a direct confrontation and, this form being closely related to
 a lattice SD equation which is just a mathematical identity, we are in a good
 position to trace back any discrepancy.

 In the
next subsection we present the continuum limit derivation  of the  Green
function formulation of the ghost SD, as well as its lattice derivation.

\subsubSection{Continuum limit case}
For a given gauge field configuration $\mathcal{A_\mu} = A^a_\mu t^a$, the 
Faddeev-Popov operator in covariant gauges is given by
\begin{eqnarray}\label{FP}
M(x,y)_{\text{1conf}} = (\partial_{\mu}D_{\mu})\delta(x-y) = \left( \Delta + i g_0 \partial_\mu\mathcal{A_\mu} 
\right) \delta^{4}(x-y) 
\equiv {F_{\text{1conf}}^{(2)}}^{-1} (\mathcal{A},x,y),
\end{eqnarray}
and it is equal to the inverse of the ghost correlator in the background of the  gluon field $\mathcal{A}$, $F_{\text{1conf}}^{(2)}$. 
The subscript means here that the equation is valid for any given gauge
configuration. This can be written 
\begin{eqnarray}
\delta^{4}(x-y) \equiv  M_{\text{1conf}}(x,z) F_{\text{1conf}}^{(2)}(\mathcal{A},z,y),
\end{eqnarray}
where a summation on $z$ is understood. Expanding $M$ according to 
(\ref{FP}) we get
\begin{eqnarray}\label{26}
\delta(x-y) = \Delta(x,z)F_{\text{1conf}}^{(2)} (z,y) +  i g_0\,\partial^{(x)}_\mu \Big( \mathcal{A_\mu}(x)   F_{\text{1conf}}^{(2)}(\mathcal{A},x,y)\Big),
\end{eqnarray}
 valid for \emph{any} gauge field configuration. 
 Performing the path integral one gets the mean value on gauge configurations 
\begin{eqnarray}\label{GSD} \delta(x-y) = \Delta(x,z)\langle F_{\text{1conf}}^{(2)} (z,y)\rangle +  i g_0 
\partial^{(x)}_\mu \langle\mathcal{A_\mu}(x)   F_{\text{1conf}}^{(2)}(\mathcal{A},x,y)\rangle . \end{eqnarray}
Of course $\langle F_{\text{1conf}}^{(2)} (z,y)\rangle$ is nothing else but the ghost propagator defined in equation~(\ref{dressghost}).

The averages $\langle F^{(2)}(x,y)\rangle$ and $
\partial^{(x)}_\mu\langle\mathcal{A_\mu}(x)   F^{(2)}(x,y)\rangle$ are  invariant 
under translations so that one can replace 
the derivative $\partial^{(x)}_\mu \to - \partial^{(y)}_\mu$. 
We take $x=0$ and perform the Fourier transform on the $y$ variable (note that there is no tilde on $\mathcal{A_\mu}(0)$ )
\begin{eqnarray}
\begin{split}
 1= - p^2 \tilde{F}^2(p)  -g_0 p_\mu \langle \mathcal{A_\mu}(0) 
 \tilde{F}_{\text{1conf}}^{(2)}(\mathcal{A},p)\rangle.
\end{split}
\end{eqnarray}
Finally we get 
\begin{eqnarray}
\label{SD_continum_Green}
 F(p^2)=  1 + g_0 \frac{p_\mu}{N_c^2 -1} f^{abc}\langle  A^c_\mu(0)  \tilde{F}_{\text{1conf}}^{(2)ba}(\mathcal{A},p)\rangle .
\end{eqnarray}
One caveat is in order
here. Eq. ~(\ref{SD}) implies an ultraviolet divergent integral which is matched
by renormalization constants. This divergence is of course also present  in eq.
(\ref{GSD}) via the local product of operators at $x$. In section
\ref{revisiting} this divergence was canceled by subtracting two terms as is
apparent in  eq. ~(\ref{k1k2}). In the following this divergence is regularised
by  the lattice cut-off. To perform the connection with the discussion in 
section \ref{revisiting} it will be necessary to perform an analogous
subtraction  when using the form (\ref{SD_continum_Green}).

\subsubSection{Lattice case}
 We now repeat the same steps as in the preceding
paragraph for the lattice version of the  Faddeev-Popov operator 
\begin{equation}
\label{FP_lattice}
\begin{split}
M_{xy}^{ab} & \equiv {F_{\text{1conf}}^{(2)}(U,x,y)}^ {-1}  =  \sum_\mu 
\Big[
S_\mu^{ab}(x) 
\left( 
        \delta_{x,y} - \delta_{y,x+\emu}
\right)
-
S_\mu^{ab}(x-\emu) 
\left( 
        \delta_{y,x-\emu} - \delta_{y,x}
\right)
+ \\ &
+\frac{1}{2} f^{abc}
\left[ 
A_\mu^c(x) \delta_{y,x+\emu} - A_\mu^c(x-\emu) \delta_{y,x-\emu}
\right]
\Big],
\end{split}
\end{equation}
where
\begin{equation}
\label{S_and_A}
\begin{array}{l}
S^{ab}_\mu(x) = - \frac{1}{2} \tr \left[ \{t^a, t^b\}\left(U_\mu(x) + 
U^\dagger_\mu(x) \right)\right]
\\
A_\mu(x) = \frac{U_\mu(x) - U^\dagger_\mu(x)}{2}-\frac{1}{N}\tr{\frac{U_\mu(x) 
- U^\dagger_\mu(x)}{2}},
\end{array}
\end{equation}
in which $U_\mu(x)$ %
denotes a standard
 link variable\footnote{Note that the definition of $A_\mu$ given in (\ref{S_and_A}) differs from the na\"\i ve one by a factor $i g_0$. This is the reason for the presence of $i$ and the absence of $g_0$ in eq.~(\ref{SD_L1_av}) as compared to eq.~(\ref{SD1}).}, and $\emu$ is a unitary vector in direction $\mu$. 
We define

\begin{equation}\label{DU}
\begin{array}{l}
\Delta_U = \sum_\mu  \left(S_\mu^{ab}(x) 
\left( 
        \delta_{x,y} - \delta_{y,x+\emu}
\right)
-
S_\mu^{ab}(x-\emu) 
\left( 
        \delta_{y,x-\emu} - \delta_{y,x}
\right)
\right).
\end{array}
\end{equation}

\noindent The appearance of $\Delta_U$  as the appropriate discretisation of the usual Laplacian operator $\Delta$  is dictated by the gauge invariance of the original Yang-Mills action, which imposes that the standard $\nabla$ operator be replaced by its \emph{covariant} version and by the specific  form -- $\mathfrak{Re}\tr (\sum_{\text{links}} U^g)$ -- of the functional to be minimized in order  to fix the Landau gauge.
 
\noindent Then, multiplying  $M_{xy}^{ab} $ by $F^{(2)}(x,y)$
from the right,
one obtains :
\begin{equation}
\label{SD_L1}
\begin{split}
 & \frac{1}{N_c^2 -1} \tr \Delta_U(y,z) F_{\text{1conf}}^{(2)}(U;z,u)   = \delta_{y,u} -
\\ &   - 
 \frac{f^{abc}}{2(N_c^2 -1)} 
\left[ 
A_\mu^c(y) F_{\text{1conf}}^{(2)ba}(U;y+\emu,u)  - A_\mu^c(y-\emu) F_{\text{1conf}}^{(2)ba}(U;y-\emu,u)
\right]
,
\end{split}
\end{equation}
This is an exact mathematical identity for each gauge configuration which must actually be fulfilled by our lattice data since our  $F_{\text{1conf}}^{(2)}$ are computed by means of an explicit inversion of the Faddeev-Popov operator. From this fact results an important feature which we wish to stress : 
\emph {since eq.~\ref{SD_L1} is valid in any configuration, its consequences are free of  any ambiguity originating from the presence of Gribov copies}.
Upon averaging over the configurations one gets

\begin{equation}
\label{SD_L1_av}
\begin{split}
 & \frac{1}{N_c^2 -1} \tr \langle\Delta_U(y,z) F_{\text{1conf}}^{(2)}(z,u)\rangle  = \delta_{y,u} -
\\ &   - 
 \frac{f^{abc}}{2(N_c^2 -1)} 
\left[ 
\langle A_\mu^c(y) F_{\text{1conf}}^{(2)ba}(U,y+\emu,u)  - A_\mu^c(y-\emu) F_{\text{1conf}}^{(2)ba}(U,y-\emu,u)
\rangle\right]
\end{split}
\end{equation}
Of course, this averaging  procedure depends on the way chosen to  treat  Gribov's problem : the particular set of configurations over which it is performed depends on the prescription which is adopted (choice of any local minimum of $A^2$,  restriction to  the fundamental modular region...) and, consequently, the Green functions will vary  but \emph{they will in any case satisfy the above equation.}
Like in the continuum case, after setting $y$ to zero, a Fourier transformation with respect to $u$ gives:
\begin{equation}
\label{SD_L1_av2}
\begin{split}
 & \frac{1}{N_c^2 -1} \tr \sum_u e^{i p\cdot u} \langle\Delta_U(0,z) F_{\text{1conf}}^{(2)}(U,z,u) \rangle  = 1 - 
\\ &   - i \sin(p_\mu)
 \frac{f^{abc}}{(N_c^2 -1)} 
 \langle A_\mu^c(0) \tilde{F}_{\text{1conf}}^{(2)ba}(U,p) \rangle
\end{split}
\end{equation}

Note that although eqs.~(\ref{SD_L1}) and (\ref{SD_L1_av})  have to be exactly verified by  lattice data eq.~(\ref{SD_L1_av2}) does only approximately (within statistical errors) since it relies on translational invariance, which  could be guaranteed only if we used an infinite number of configurations.

Eq.~(\ref{SD_L1_av2}) is a discretised version of (\ref{SD_continum_Green}). Therefore any lattice correlator,  satisfying (\ref{SD_L1_av2}), should also satisfy   (\ref{SD_continum_Green}) \emph{up to non-zero lattice spacing effects}. Among the various  sources for such effects 
 the use of the specific $\Delta_U$ discretisation of the Laplacian operator in the l.h.s. deserves 
some comments. The  gauge fields present in the $S_\mu^{ab}(x)$ terms
 in eq.~(\ref{DU}) generate in  $\Delta_U$ the 
so-called ``tadpole'' diagrams such as in fig. \ref{tadpole}.
 \begin{figure}[h]
\begin{center}
\psfig{figure=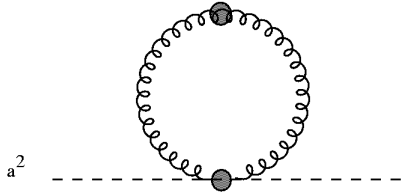,height=3cm}
\end{center}
\caption{Example of the terms in the Schwinger-Dyson equation on the lattice.}
\label{tadpole}
\end{figure}
According to
the philosophy developed by Lepage and Mackenzie in ref.~( \cite{Mac}) the tadpole contribution can be estimated by a mean field method. Using the average plaquette  $P$ 
(for $\beta=6.0\, P\simeq 0.5937$) one predicts 
a tadpole correction factor $\propto P^{-(1/4)} \simeq 1.14$. These terms disappear in the continuum limit  but they do so only very slowly : the
tadpole  corrections (1 - plaquette) vanish  only as an inverse  logarithm with the
lattice spacing. This is to be contrasted with the corrections arising in the r.h.s which are expected to be of order $a^2$.

\subsubSection{Checking the validity of the SD equation on the lattice}\label{DeltaU}
Since, as we have just mentionned,   eqs.~(\ref{SD_L1}) and (\ref{SD_L1_av}) are  mathematical identities, there is in principle little --if any-- to learn from a verification of eq.~(\ref{SD_L1_av2}), except for the verification that our configurations are actually in the Landau gauge.  On the other hand one thing we wish to be reassured about is  the possible role of lattice artefacts in the discrepancies we have noticed.

We begin with a comparison of the continuum r.h.s of eq.~(\ref{SD_continum_Green}) with the l.h.s. of~(\ref{SD_L1_av2}).
Both sides  are plotted in 
fig.~(\ref{WITH_DELTA_U}). 

%

The agreement is  impressive. Should we on the contrary use eq.~(\ref{SD_continum_Green}) itself we  observe a clear disagreement between the two sides of the equation.
What we thus learn is that the major part of the discretization artefact
comes from  $\Delta_U$. In fact, our lattice 
data show that $\Delta_U \simeq \Delta/1.16$ {\it almost independently of the
momentum}, see fig. \ref{SD_cont_lattice}. This is in good agreement with
the  correction factor of 1.14 obtained from Lepage-Mackenzie's mechanism \cite{Mac}. To conclude the tadpole effect explains 
almost all of the discrepancy observed when trying to verify (\ref{SD_continum_Green}).
One can also understand why this discretisation artefact is so large : this is due to the slow logarithmic vanishing of the 
tadpole  corrections. 

\begin{figure}[h]
\begin{center}
\psfig{figure=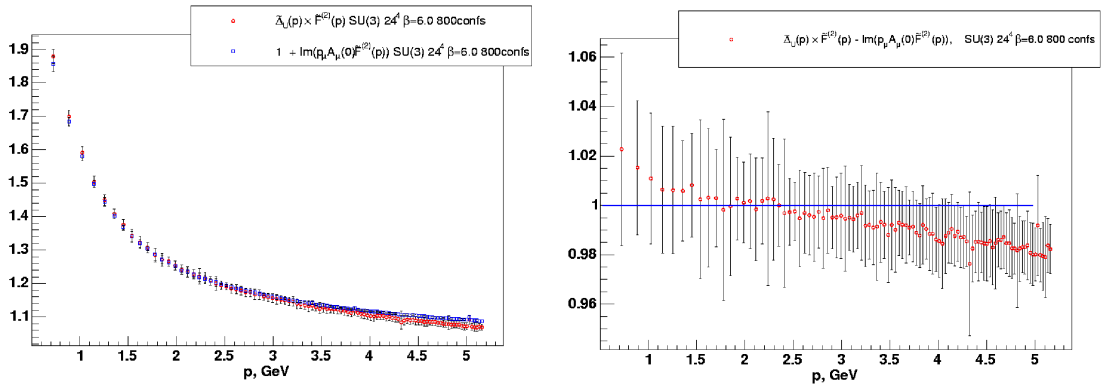, height=15cm, angle=-90}
\end{center}
\caption{Check of the validity of the SD equation on the lattice. 
\underline{Left}~: ~$\frac{1}{N_c^2 -1} \langle\Delta_U(p) \tr \,\tilde{F}_{\text{1conf}}^{(2)} (p)\rangle$ (circles) is plotted vs. $1 + g_0 \frac{p_\mu}{N_c^2 -1} f^{abc}\langle A^c_\mu(0)  \tilde{F}_{\text{1conf}}^{(2)ba}(\mathcal{A},p)\rangle$ (squares). 
\underline{Right} : $\frac{1}{N_c^2 -1} \langle\Delta_U(p) \tr \,\tilde{F}^{(2)} (p)\rangle - g_0 \frac{p_\mu}{N_c^2 -1} f^{abc}\langle A^c_\mu(0)  \tilde{F}_{\text{1conf}}^{(2)ba}(\mathcal{A},p)\rangle$ is compared to $1$.}
\label{WITH_DELTA_U}
\end{figure}

It results from this discussion that the lattice artefacts cannot be blamed 
for the violation of the relation   \Ral   observed   in fig. ~(\ref{Both_F2G}) : the tadpole effect has been seen to produce 
a corrective factor almost constant in $p$, thus unable 
to explain an error in the power behaviour.

%
\begin{figure}[h]
\begin{center}
\psfig{figure=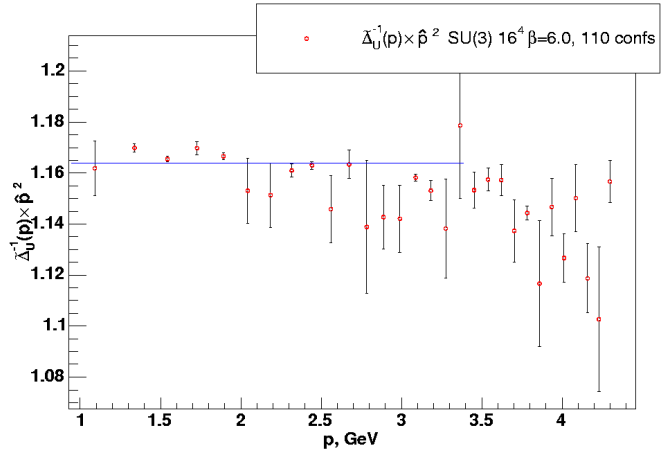, height=9cm, angle=-90}
\end{center}
\caption{Lattice computation of $\Delta/\Delta_U$ in Fourier space  as a function of the momentum. The statistics is poor for technical reasons. Note also that the region  above $\pi/2\,(\sim 3\,  \text{GeV})$ is affected by strong discretisation effects.}
\label{SD_cont_lattice}
\end{figure}
\subsubSection{Concluding remarks}\label{concluSD}

We want to emphasize at this stage that our lattice data both \emph{satisfy} the properly discretised SD equation and \emph{violate} the relation \Ral. 
The most likely way out we can think of is that the
hypothesis $\alpha_\Gamma=0$  is not verified,  i.e. that \underline{$H_1$   is singular when all}  \underline{momenta are small}. One  more argument in favour of this explanation is provided by a direct numerical examination which shows that, in the eventuality of  neglecting the momentum dependence of $H_1$, the SD equation is satisfied in the UV but badly violated in the IR. The results of this comparison are given in fig.~\ref{test_trunc_SD}   . The details of the method can be found in the appendix.
%
\begin{figure}[h]
\begin{center}
\psfig{figure=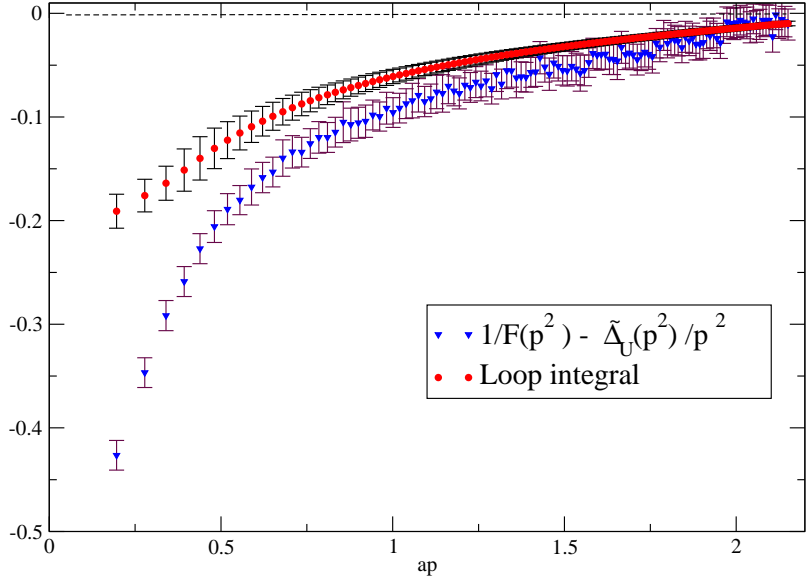, height=13cm, angle=-90}
\end{center}
\caption{Testing the truncated ghost SD equation on the lattice. The data correspond to a
$SU(3)$ simulation at $\beta=6.4$ and $V=32^4$,  ($a^{-1}=3.66$ GeV).}
\label{test_trunc_SD}
\end{figure}

 The possibility of a singular behaviour of $H_1$ in the infrared has already been considered by various authors, for example in  ref.~\cite{Ellwanger} in the framework of exact renormalization group equations.  It is known from  perturbation 
theory (\cite{Taylor}), that the sum  $H_1 + H_2$ (cf eq.~(\ref{h1h2})) is exactly  equal to 1 at $k = 0$ (this has been checked explicitly in  perturbation theory  up to fourth
 order in ref.~\cite{Chetyrkin:2000dq}). This is the argument which is  usually called for when advocating $\alpha_\Gamma=0$. However 
 it does not prevent the contributions of the scalar function $b$   to $H_1$ and $H_2$, which cancel out in the sum, from being singular in this limit.  
 Actually, from  dimensional considerations,  one concludes  that $b$ must be of dimension $-2$. At $k =0$ the only dimensional quantity involved
  (at   the perturbative level) is $q$, which means $b(q,0;-q) \propto 1/q^2$.  This singularity is removed by the kinematical factor in front of $b$ in
   $H_1$ and $H_2$, but this would no longer be the case for $k \ne 0$ if one had more generally  $b(q,k;-q-k) \propto 1/q^2$. \emph{In any case our results appear
    to plead in favour  of  
 a divergent ghost-ghost-gluon vertex in the infrared domain}.


\Section{Slavnov-Taylor identity and the infrared behaviour of the gluon propagator}\label{STID}

Another non-perturbative relation that can be exploited is the Slavnov-Taylor
identity. It can be used to constrain (under some  hypotheses) the infrared exponent
for the gluon dressing function.  In the preceding section  we have explored the consequences of the very strong assumption that $H_1$  is regular when all its arguments go  to zero and we have shown that this assumption is not tenable when the lattice data are taken into account.   We shall now make  the weaker hypothesis that the scalar factors present in the decomposition~(\ref{vertghost}) of the ghost-ghost-gluon vertex are regular when one of their arguments go to zero while the others are kept finite and
 exploit the Slavnov-Taylor identity under this assumption to derive constraints on $\alpha_F$ and $\alpha_G$.

The Slavnov-Taylor (\cite{Taylor}) identity for the three-gluon function
reads
\begin{equation}
\label{STid}
\begin{split}
p^\lambda\Gamma_{\lambda \mu \nu} (p, q, r) & =
\frac{F(p^2)}{G(r^2)} (\delta_{\lambda\nu} r^2 - r_\lambda r_\nu) \widetilde{\Gamma}_{\lambda\mu}(r,p;q) - 
\\ & -
\frac{F(p^2)}{G(q^2)} (\delta_{\lambda\mu} q^2 - q_\lambda q_\mu) \widetilde{\Gamma}_{\lambda\nu}(q,p;r).
\end{split}
\end{equation}
One then takes the limit $r \rightarrow 0$ while  keeping $q$ and $p$ finite. The tensor structure of $\widetilde{\Gamma}_{\mu\nu}(p,k;q)$
 has been recalled in eq.~(\ref{vertghost}). 
 We adopt the following notations for particular kinematic configurations:
\begin{equation}
\begin{array}{l}
a_3(p^2) = a(-p,p; 0 ) \\
a_1(p^2) = a(0, -p; p), \quad b_1(p^2) = b(0, -p; p), \quad d_1(p^2) = d(0, -p; p).
\end{array}
\end{equation}
In the present case of one zero momentum the three-gluon vertex may be parameterised in the general way as (\cite{Chetyrkin:2000dq})
\begin{equation}
\Gamma_{\mu\nu\rho}(p,-p,0) = 
\left( 2\delta_{\mu\nu}p_\rho - \delta_{\mu\rho}p_\nu - \delta_{\rho\nu}p_\mu\right) T_1(p^2) - 
\left(  \delta_{\mu\nu} - \frac{p_\mu p_\nu}{p^2} \right) p_{\rho}T_2(p^2) +
p_\mu p_\nu p_\rho T_3 (p^2).
\end{equation}

\noindent Now, exhibiting the dominant part of each term of eq.(\ref{STid}) we obtain:
\begin{eqnarray}
\label{STlimr0}
& T_1(q^2)(q_\mu q_\nu - q^2 \delta_{\mu\nu})+q^2 T_3(q^2)q_\mu q_\nu +\eta_{1\mu\nu}(q,r)= \nonumber \\
& \frac{F((q+r)^2)}{G(r^2)}\Big[(a_1(q^2)+r_1(q,r))(\delta_{\mu\nu}r^2-r_\mu r_\nu)1+
 +(b_1(q^2)+r_2(q,r))
q_\mu(r^2 q_\nu-q.r r_\nu) \nonumber \\&+(b_1(q^2)+d_1(q^2)+r_3(q,r))r_\mu(r^2 q_\nu - q.r r_\nu)\Big] + \nonumber \\
& +\frac{F((q+r)^2)}{G(q^2)}\Big[a_3(q^2)(q_\mu q_\nu - q^2 \delta_{\mu\nu}) +\eta_{2\mu\nu}(q,r)\Big]
\end{eqnarray}
where $r_{1,2,3}$ and $\eta_{1,2}$ verify
\begin{eqnarray}
\begin{split}
& \lim_{r\rightarrow 0 }r_1(q,r) = \lim_{r\rightarrow 0 }r_2(q,r) = \lim_{r\rightarrow 0 }r_3(q,r) = 0 \\
& \lim_{r\rightarrow 0 }\eta_{1\mu\nu}(q,r) = \lim_{r\rightarrow 0 } \eta_{2\mu\nu}(q,r) = 0 \\
\end{split}
\end{eqnarray}
Identifying the leading terms of the scalar factors multiplying the tensors $q_\mu q_\nu$ and
$q_\mu q_\nu - q^2 \delta_{\mu\nu}$ we obtain the usual relations (\cite{Chetyrkin:2000dq}):
\begin{equation}
\label{WIhab}
\begin{array}{l}
T_1(q^2) = \frac{F(q^2)}{G(q^2)}a_3(q^2) \\
T_3(q^2) = 0.
\end{array}
\end{equation}
Using these relations we see that Eq.(\ref{STlimr0}) implies:
\begin{equation}
\lim_{r\rightarrow 0}\frac{F(p^2)}{G(r^2)}\Big[a_1(q^2)(r^2\delta_{\mu\nu}-r_\mu r_\nu)+b_1(q^2) (r^2 q_\mu q_\nu -r.qq_\mu r_\nu)\Big] = 0
\end{equation}
Thus one sees that if $a_1(q^2)\neq 0$ or $b_1 \neq 0$ (and,   indeed, one knows from  
perturbation theory that at large momenta $a_1$ = 1, cf.~\cite{Taylor,Chetyrkin:2000dq}) (\ref{STid}) can only be compatible with the parameterisation~(\ref{param}) if 
\begin{equation}
\alpha_G < 1. 
\end{equation}

We can also, instead of  letting $r \to 0$,   take the limit $p\rightarrow 0$ of Eq.(\ref{STid}) as is done in \cite{Chetyrkin:2000dq}.
The dominant part of the l.h.s. of (\ref{STid}) is :
\begin{equation}
(2\delta_{\mu\nu}p.q-p_\mu q_\nu-p_\nu q_\mu)T_1(q^2)-(\delta_{\mu\nu}-\frac{q_\mu q_\nu}{q^2})p.qT_2(q^2)
+p.q q_\mu q_\nu T_3(q^2) \nonumber
\end{equation}
The r.h.s. is the product of $F(p^2)$ with an expression of at least first order in p.  $T_1$ and
$T_2$ being different from zero we can conclude in this case that $\alpha_F \le 0$. \\

Let us repeat here that all these considerations are valid only if all scalar factors
of the ghost-ghost-gluon and three-gluons vertices are regular functions when one momentum goes to zero while the others remain finite.
\underline{Under those hypotheses} one obtains important constraints on the gluon and ghost propagators - namely that they are divergent in the
zero momentum limit.
 
\subSection{Lattice  results}
The results for $\alpha_G$ and $\alpha_F$ from lattice simulations are presented
in figs.~ (\ref{Both_F2G}) and (\ref{plotSU2}).  We present in Table~\ref{tablalpha} the values of the coefficients for a fit of the form\footnote{ A term of the form $\mu q^2$ is clearly needed in order to describe a situation like  the one in fig.~(\ref{plotSU2} left) where $G^{(2)}(p^2)$ seems to go  to a finite limit when $p$ goes to zero} $(q^2)^\alpha (\lambda + \mu q^2)$.
\begin{table}[h]
\begin{center}
\begin{tabular}{|c|c|c||c|c|}
\hline
Group&Volume&$\beta$&$\alpha_G$&$\alpha_F$\\
\hline
SU(2)&$48^4$&$2.3$&$1.004\pm 0.015$&$-0.087\pm0.015$\\
\hline
SU(2)&$32^4$&$2.3$&$0.968\pm0 .011$&$-0.109\pm0.014$\\
\hline
SU(3)&$32^4$&$5.75$&$0.864\pm 0.016$&$-0.153\pm0.022$\\
\hline
\end{tabular}
\caption{Summary of the fitting parameters for the F and G functions}\label{tablalpha}
\end{center}
\label{default}
\end{table}%

%
%

\begin{figure}[h]
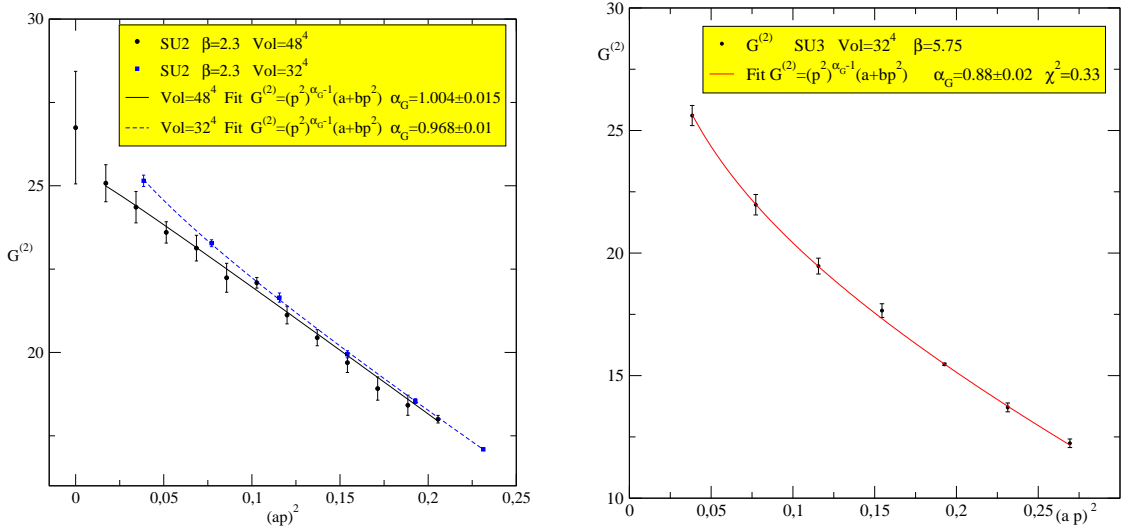

\begin{center}
\hspace{-.5cm}\begin{tabular}{lcr}
\includegraphics[width=7.cm]{G2_SU2_48_32.eps}&&\includegraphics[width=7.cm]{G2_SU3.eps}\\
\end{tabular}
\end{center}
\caption{$G^{(2)}(p^2)$ from lattice simulation for  $SU(2)$ (left) and $SU(3)$ (right) . $\beta_{\text{SU(2)}}=2.3$ and $\beta_{\text{SU(3)}}=5.75$. the  volumes are $32^4$ and$48^4$ for $SU(2)$ and $32^4$  for $SU(3)$.} 
\label{plotSU2}
\end{figure}

 The fits have been performed without using the point $p^2$ = 0 even in the case of the  gluon propagator where it is known.  Its inclusion would have \emph{forced} $\alpha_G$ to be equal to 1.  In any case one sees on fig.~\ref{plotSU2} that the point at $p^2 = 0$ available in the $SU(2), 32^4$ case is compatible with the fit.

 For SU(2) and the larger volume the value obtained for $\alpha_G$ is  compatible with 1. 
 The situation is less clear for SU(3), but in this case data with the larger volume ($48^4$) are lacking.  Moreover we have to take into account our experience from previous studies of the gluon propagator where
  we have always observed that the gluon propagator goes continuously to a finite limit in the infrared region.
  A very detailed study of the gluon dressing function and specially of its  volume dependence at $k=0$  has been performed by Bonnet et al. 
 (cf ref.~\cite{Adelaide}).   This study shows that a value $\alpha_G = 1$ is compatible with the data (the dressing function shows no signal of discontinuity in the neighbourhood of zero) and that no pathology  shows up as the volume goes to infinity.
 
 We conclude that lattice data seem to contradict Zwanziger's result ($G^{(2)}(0)\, =\, 0 $), \cite{Zwanziger:1991gz} and most probably also  the predictions derived from  the Slavnov-Taylor identity ($G^{(2)}(0)$ infinite).  Like in the preceding section a possible way out of this contradiction could  consist in dropping the regularity assumptions which have been made in the course of the proof.

\Section{Discussion and Conclusions}
\subSection{Discussion of the validity of SD and ST equations in the infrared}

Schwinger-Dyson equations and Slavnov-Taylor identities are valid
non-pertur\-ba\-ti\-ve\-ly. However some care is needed mainly due to
Gribov ambiguities in the gauge fixing procedure.  
One needs   a well defined QCD partition function for a given gauge (in the
following we concentrate on Landau gauge). Many
different  non-perturbative prescriptions for the quantisation of non-abelian
theories have been suggested : integration only
over the absolute minima of the $\int A^2$ functional  (Gribov fundamental
modular region)~\cite{Zwanziger:1991gz}, summing of copies with signed
Faddeev-Popov  determinant~\cite{Baulieu:1983tg},
stochastic quantisation~\cite{Zinn-Justin:1987ux},
etc. All these prescriptions correspond to different valid gauge fixing
procedures.

In lattice numerical simulations two main methods are of practical use:
The algorithm which minimises the functional is stopped at its first solution,
which is a local minimum, or one takes the smallest local minimum among a given
number of trials on the same gauge orbit.  One is sure to be inside the Gribov
region, never to be inside the fundamental modular region.

The question is whether SD and ST relations are valid in these gauge fixing
schemes. It is argued in (\cite{Zwanziger:2001kw}) that the
 Schwinger - Dyson equations are valid under different quantisation
prescriptions provided that the Faddeev-Popov determinant vanishes on
the boundary of the integration domain. However, the partition functions will
differ and hence, the Green functions will be in general different solutions of the
the SD equations.

The Slavnov-Taylor identities
may be derived from the QCD partition function  using the gauge invariance
of the action and stating that the gauge fixed path integral
is invariant after a change of variables which corresponds to a gauge
transformation. It is then clear that the Slavnov-Taylor identities
remain valid whichever gauge fixing procedure has been followed.
However this proof has to be taken with care in the presence of
singularities of the partition function or of Green functions.

One more general comment is in order. As we have already stated, no Green
function with a vanishing ghost  momentum can be defined on the lattice since
the zero modes of the Faddeev-Popov operator are discarded. For the same reason
no source term for the zero mode ghost field is allowed. More generally we do
not know of any non perturbative way to define  such  Green functions. We
cannot prove that this means a divergence of the Green  functions with one
ghost momentum going to zero, but we can  suspect that it is the  case as
lattice indicates for the ghost propagator~\footnote{Notice that we cannot
claim anything  about the behaviour of vertex functions when one  ghost
momentum goes to zero.}. Indeed the close-to-zero modes of the  Faddeev-Popov
operator are strongly influenced by the Gribov horizon and, for sure, very
different from any perturbative result.  This casts also some doubt about the
use of ST identities or SD equations in the case of a zero ghost momentum.

\subSection{Conclusions}
We have tried in this work to put together various inputs in order to
clarify our understanding of the infrared behaviour of the pure Yang-Mills Green functions.  Our findings can be summarised as follows :

\begin{enumerate}
\item The lattice Green functions contradict the common {\sl lore} according to which $2 \alpha_F + \alpha_G = 0$. The present situation  is that  $2 \alpha_F + \alpha_G > 0$, i.e., the product $F^2(k^2)\, G(k^2)$ tends to $0$ for $k \to 0$. From what we observe concerning the evolution of the curves with the size of the lattice,  it is difficult to imagine how a further increase  of the volume could eventually revert this tendency.
 
     \item  The result $2 \alpha_F + \alpha_G = 0$ (\Ral)  which is contradicted by lattice data, is usually claimed to be derived from the ghost SD equation.  Our results seem to cast some doubt on the validity of these derivations, based on the assumption of a trivial ghost-ghost-gluon vertex (``na\"\i ve approximation''). Indeed, we do verify that the \emph{properly discretised} SD ghost equation on the lattice is {well satisfied}. We conclude that
 the lattice data seem to prove that this ``na\"\i ve approximation'' is invalid, and that there exists in the ghost-ghost-gluon vertex a non-perturbative infrared singularity of the form $(k^2)^{\alpha_\Gamma}$ which has been neglected in the standard analysis and which leads to the replacement of \Ral by $2 \alpha_F + \alpha_G  +\alpha_\Gamma = 0$ ($\alpha_\Gamma<0$).

\item Regarding the ghost correlator it results from its very definition that it cannot be defined at zero momentum.  Its lattice values at low momentum appear to be in favour of a divergent dressing function ($\alpha_F < 0$), as is also suggested by the above theoretical arguments. However the divergence is much slower ($\alpha_F\,\in[-.15,\,-.1]$) than what is usually reported. This is in agreement with the conclusions one can draw from the Slavnov Taylor identity.

\item In relation with points 2) and 3) above it is worth insisting on the fact that the Schwinger -Dyson equation by itself is not sufficient to determine the behaviour of the ghost-propagator and of the ghost-ghost-gluon vertex. Different treatments of the Gribov copies  lead to different infrared solutions (se also ref~\cite{Bakeev:2003rr} in this respect), \emph{all of which fulfill the SD equation}.  

\item  As for the gluon propagator the situation is much less clear. Three sources of information are available and give contradictory results.
\begin{itemize}
\item[-] The Slavnov Taylor identity, supplemented by  regularity assumptions for the ghost-ghost-gluon vertex functions, points towards a \emph{divergent} infrared behaviour.
\item[-] Our lattice data  indicate a  finite limit when the momentum goes to zero. This  trend is very clear in the $SU(2)$ case although less compelling for  $SU(3)$. The fit has been performed by excluding the point at $ p=0$ but the  latter, when known, is compatible with the extrapolated value. These results agree with the ones of our previous studies as well as with the findings of the other lattice groups who have studied this matter (\cite{Adelaide}) which include the point $ p=0$ and impose therefore $\alpha_G = 1$.
\item[-] Zwanziger's result \cite{Zwanziger:1991gz} states that  the gluon propagator vanishes at k=0 but a fully satisfactory proof of a \emph{continuous} vanishing as $k \to 0$ is still lacking.  
\end{itemize}
   We are unable for the moment to settle this point in a totally unambiguous way.
\end{enumerate} 

 This set of conclusions raises some questions. First of all further studies are still under way in order to fix the issues related to point 4). As to point 2), a direct lattice study of the ghost-ghost-gluon vertex at low momenta is desirable, although difficult. Such a study has recently been performed for  $SU(2)$ in a specific kinematical situation (zero-momentum gluon) (\cite{CucchiMendes}). The precise relevance of this special case to  the points we have considered remains to be clarified.
 
 \section*{Note added} After the completion of this work we realised that the particular situation ($\alpha_F = 0$)  which  we have mentionned in the fourth column of table \ref{tabledescas}  but not fully investigated might  provide a good agreement with the lattice data while complying with the constraints stemming from the SD equation. This possibility is discussed in a further publication  (\cite{Finite:2006}) .  Let us  stress that  this solution too is compatible with the non-vanishing of $2 \alpha_F + \alpha_G$.
 
\Section{Acknowledgements}We wish to thank N. Wschebor and U. Ellwanger for illuminating discussions.

\section{Appendix: Testing the naive approximation of the ghost SD equation}

The simplest approximation of the ghost SD  equation~(\ref{SD1}) corresponds to the case $H_1(q,k)=1\,\forall q,k$:
\begin{equation}
\label{trunc_SD_cont}
\frac{1}{F(k)} = 1 + \frac{g_0^2 N_c}{k^2} \int \frac{d^4 q}{(2\pi)^4} 
\Big(\frac{F(q^2)G((k-q)^2)}{q^2 (k-q)^2} 
\frac{(k\cdot q)^2 - k^2 q^2}{(q-k)^2}
\Big]\Big)
\end{equation}
Strictly speaking this equation, written in this way, is meaningless since it involves UV divergent quantities but a corresponding meaningful renormalized version can be given (see the caveats about eq~(\ref{SD}) in subsection ~(\ref{revisiting})).
We want to check whether lattice propagators satisfy it. According to  perturbation theory,
it should be true at large $k$. Lattice propagators are discrete functions, and thus one has to handle the problem 
of the numerical evaluation of the loop integral $I$ in (\ref{trunc_SD_cont}). Let us express the integrand 
in terms of $q^2$ and $(k-q)^2$
\begin{equation}
\label{I_is}
I =\int \frac{d^4 q}{(2\pi)^4} 
\frac{F^2(q)G(k-q)}{q^2 (k-q)^2 } 
\Big[ \frac{(k-q)^2}{4} + \frac{(k^2)^2+(q^2)^2 - 2k^2 q^2}{4(k-q)^2} 
-\frac{q^2+k^2}{2} \Big].
\end{equation}
Then we write
$$
I=I_1+I_2+I_3+I_4+I_5+I_6,
$$
each $I_{i}$ corresponds to one term in (\ref{I_is}). 

All these integrals have the form 
$$
I_i = C_i(k) \int \frac{d^4 q}{(2\pi)^4} f_i (q) h_i(k-q).
$$
The convolution in the r.h.s. is just the Fourier transform of the product at the same point
in  configuration space:
$$
\int \frac{d^4 q}{(2\pi)^4} f_i (q) h_i(k-q) = \text{F}_{+}
\Big(  \text{F}_{-}(f_i)[x] \text{F}_{-}(h_i)[x]  \Big)(k),
$$
where $\text{F}_{-}(\hat{f})(x)$ is an inverse and $\text{F}_{+}(f)(k)$ a direct Fourier transform. 
Thus, in order to calculate the integral $I$ from discrete lattice propagators one proceeds as follows:
\begin{enumerate}
\item calculate $\{f_i\}(p)$  and $\{h_i\}(p)$ as functions of $F(p), G(p), p^2$ for all $i$
\item apply the inverse Fourier transform $\text{F}_{-}$ to all these functions and get $f_i(x)$ and $h_i(x)$
\item compute the product at the same point $f_i(x)\cdot h_i(x)$
\item apply the direct Fourier transform $\text{F}_{+}$ to $f_i(x)\cdot h_i(x)$
\end{enumerate}
The calculation of Fourier transforms  involves a Hankel transformation which is numerically evaluated by means of a  Riemann sum
\begin{equation}
\label{HT}
f(r) = (2\pi)^{-2} \| r \|^{-2}
\sum_{i=1}^{N} J_{1}(r \rho_i ) \rho_i^{2} \frac{\hat{f}[\rho_i] + \hat{f}[\rho_{i-1}]}{2} (\rho_i-\rho_{i-1}),\qquad \rho_0 = 0.
\end{equation}
The inverse transformation is done in the similar way. In  practice, because of the lattice artefacts which become important at large $\rho$ the summation has to be restricted to $\rho < \rho_{max} \simeq 2.2 $ instead  of the ``ideal'' value $ 2 \pi$.

\paragraph{Errors}

There are three important sources of errors :
statistical Monte-Carlo errors for $F(q^2)$ and $G(q^2)$, the bias due to the 
integral discretization  and the truncation of the $\rho$-summation to values lower than $\rho_{max}$. The second one is dominated by the neglected contribution
coming from the UV cut-off of the integral (the integration is
performed on some ball $B(0,L)$ instead of $\mathbb{R}^4$).
Let us estimate the error on the Fourier transform of the product in such a case:
\begin{equation}
\begin{split}
\text{F}_{+}(f(x)g(x))(k)  = \int_{B(0,L)}d^4p d^4q \delta_{\epsilon}(k-p-q)\hat{f}(p)\hat{g}(q),
\end{split}
\end{equation}
where the $\epsilon-$approximation to delta function is $\delta_\epsilon(p) = \int_{B(0,L)}d^4x e^{i(x,p)}$.
Considering $\epsilon$ small enough we can integrate on $q$ around the point $(k-p)$,
obtaining finally:
$$
(f\star g)_L \approx (f\star g)_\infty + \text{Vol}(B(0,\epsilon(L)))\cdot(f\star \nabla g)_\infty
$$
This gives us an estimation of the error coming from the UV cut-off. As for the last source of error, it may be neglected because of the following argument : the integral is logarithmically divergent, therefore the neglected behaves as 
$\log\left(\frac{2 \pi}{L a}\right) -\log\left(\frac{\rho_{max}}{L a}\right) = \log(\frac{2 \pi}{\rho_{max}})$. Thus it remains finite as $a$ goes to zero and gets smaller and smaller as compared  to the part actually computed.

\paragraph{Results:}
We still have to face the same problem we have already encountered in section~(\ref{DeltaU}), namely that the lattice Faddeev-Popov operator involves the non trivial discretisation $\Delta_U$ of the Laplacian operator.  This is taken into account by means of the substitution of $\widetilde{\Delta_U}(p^2)/p^2$ to the ``1''  term in the l.h.s of equation~(\ref{trunc_SD_cont})
We present on (Fig. \ref{test_trunc_SD}) the result of the numerical integration 
described above. We have chosen for this purpose the data set from the simulation
with the gauge group $SU(3)$ at $\beta=6.4, V=32^4$.
One sees that the equality is achieved at large momenta, but in the infrared 
the naive approximation of the ghost SD equation fails.


\end{document}